\documentclass{PoS}

\title{Magnetic monopoles in high temperature QCD}

\ShortTitle{Magnetic monopoles in high temperature QCD }

\author{\speaker{Alessio D'Alessandro} \\
        Universita' di Genova \& INFN\\
        E-mail: \email{adales@ge.infn.it}}

\author{Massimo D'Elia\\
        Universita' di Genova \& INFN\\
        E-mail: \email{delia@ge.infn.it}}

\abstract{We study the density and spatial distribution of thermal
Abelian monopoles in the deconfined phase of SU(2) pure gauge
theory: they display non-trivial interactions with a well defined
continuum limit. The Maximal Abelian Gauge (MAG) has been chosen
to perform the Abelian projection. Questions related to the choice 
of the Abelian projection, as well as possible directions for
future studies, are discussed.} 

\FullConference{8th Conference Quark Confinement and the Hadron Spectrum \\
         September 1-6 2008\\
         Mainz, Germany}

\begin{document}

\section{Introduction}
In the dual superconductor framework of the QCD
vacuum~\cite{scmodel} Abelian magnetic monopoles may explain color
confinement, which is related to the spontaneous breaking of a
magnetic Abelian symmetry induced by monopole condensation. The
magnetic condensate in the confined phase, as well as its
disappearance at the deconfining transition, has been extensively
observed on the lattice~\cite{condensate}. Magnetically charged
particles have also been supposed to be relevant in explaining the
physical properties of the Quark Gluon Plasma phase above the
transition~\cite{magimportance,chezak}, such as the very low
viscosity and diffusion coefficient \cite{shuryakrev} and the
strongly interacting liquid-like nature~\cite{chezak,shuryakrev}.

In Ref.~\cite{chezak} the magnetic component of the Quark Gluon Plasma (QGP)
has been identified with Abelian magnetic monopoles evaporating
from the condensate and becoming thermal particles for $T > T_c$. A
definition was given for their detection on the lattice in terms of
non-trivially wrapped trajectories
in the Euclidean time direction.
In this work we apply this definition to find the physical properties
of these objects, such as their density and their
correlation functions, 
in a wide temperature range above $T_c$. 
A full account of these results is presented in Ref.~\cite{ourpaper}.


\section{The physical properties}
Simulations have been done for $SU(2)$ pure gauge for various lattice sizes
and couplings.
The physical scale has been determined as
$a(\beta) \Lambda_L = R(\beta) \lambda(\beta)$, where $R$ is the two-loop
$\beta$-function, while $\Lambda_L$ and the non-perturbative correction 
$\lambda(\beta)$ have been taken from Ref.~\cite{karsch}.

We identify monopole currents with the usual De-Grand Toussaint
construction~\cite{degrand}, after Abelian projection in the 
Maximal Abelian Gauge. The gauge fixing has been performed by
maximizing, with a standard overrelaxation algorithm, the MAG functional
\begin{equation}
F_{\rm MAG} = \sum_{\mu,x} {\rm Re}\,  \mbox{tr} \left[U_\mu(x)
  \sigma_3 U^{\dagger}_\mu(x) \,
\sigma_3\right] \label{maxfun}
\end{equation}
After that, monopole currents are defined~\cite{degrand} as $ m_\mu
= {1 \over 2 \pi} \varepsilon_{\mu\nu\rho\sigma} \hat\partial_\nu
\overline \theta_{\rho\sigma} $, where $\overline
\theta_{\rho\sigma}$ is the compactified part of the Abelian
projected plaquette $\theta_{\rho\sigma}$ and $m_\mu$ is an integer.
Being closed ($\hat\partial_\mu m_\mu = 0 $) these currents may be
trivially or non-trivially wrapped along the time direction. 
The thermal monopole density~\cite{chezak,bornyaEjiri} is given in terms of 
the temporal winding number
$N_{wrap}(m_0(\vec{x},t))$ of each time-directed current $m_0$ in
$(\vec{x},t)$ ($V_s=(L_s a)^3$ is the spatial volume):
\begin{equation}
\rho = \frac{\left< \sum_{\vec{x}} \left| N_{wrap}(m_0(\vec{x},t))
\right| \right> }{V_s}
\label{densdef}
\end{equation}

\begin{figure}[!htbp]
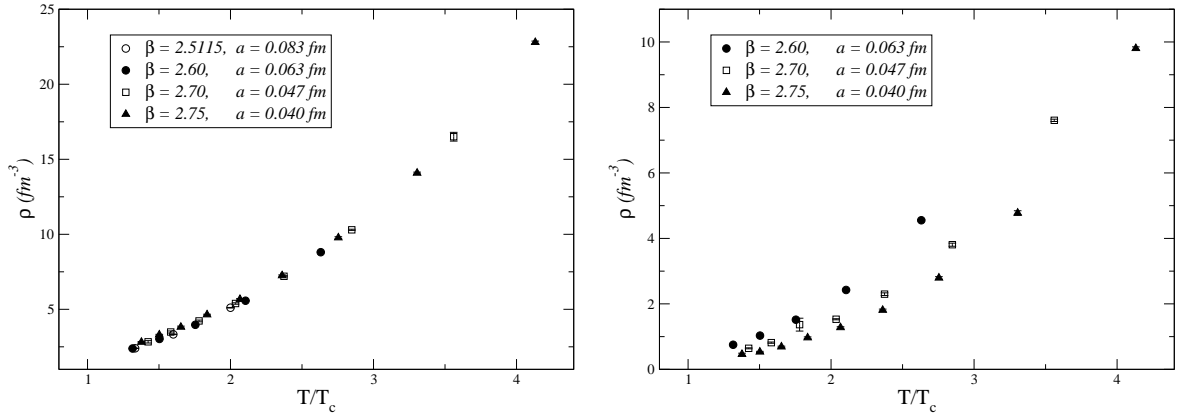

\begin{tabular}{cc}
\includegraphics[width=0.5\textwidth]{densphys.eps} & \includegraphics[width=0.5\textwidth]{densphys_LMAG.eps}
\end{tabular}
\caption{Monopole density with MAG gauge fixing from thermalized configurations [left]
and with MAG gauge fixing after Landau preconditioning [right]. Spatial lattice
sizes and $\beta$'s are respectively $(L_s,\beta)=(24,2.5115),(32,2.6),(40,2.7),(48,2.75)$ while
the temporal sizes range from $L_t=4$ to $L_t=L_s/4$.}
\label{densfig}
\end{figure}

Our results for different lattice spacings are reported in
figure~\ref{densfig} (left).
A nice scaling to the continuum limit can be observed.
The curve does not fit a simple
$\rho \propto T^3$ behavior expected for free massless particles.
A nice fit is instead obtained with \mbox{$\rho= A T^3/\log^\alpha(T/\Lambda_{eff})$}
with $A = 0.48(4)$, $T_c/\Lambda_{eff} = 2.48(3)$, $\alpha=1.89(6)$
and a $\chi^2/{\rm d.o.f.}$ of order 1. At high temperatures also $\alpha=3$, which
is the exponent expected~\cite{chezak,giovannangeli} from dimensional
reduction and perturbative considerations
($\rho\sim(g^2 T)^3$) fits well. We conclude that interactions are important
also at high temperatures.

\begin{figure}[!t]
\begin{tabular}{cc}
\includegraphics[width=0.47\textwidth]{gr_t.eps} & \includegraphics[width=0.53\textwidth]{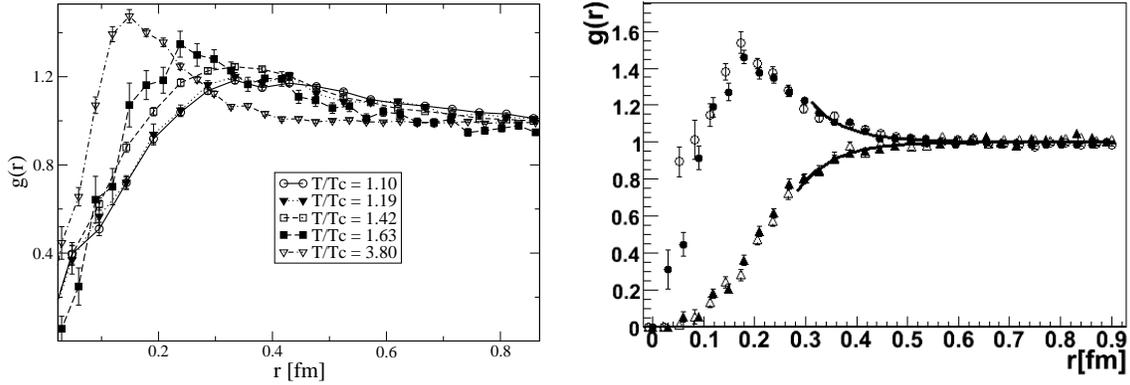}
\end{tabular}
\caption{$g_{+-}(r)$ for $\beta=2.70$ and $L_s^3\times L_t =64^3\times 8$ at different $T/T_c$  [left];
$g_{+-}(r)$ (circles) compared with $g_{++}(r)$ (triangles), for $40^3\times 5$ at $\beta=2.70$ (empty markers)
and $64^3\times 8$ at $\beta=2.86$ (full markers): both lattices correspond to the same $T/T_c\sim2.85$ [right].
}
\label{grfig}
\end{figure}

The other quantity we look at is the monopole-(anti)monopole
correlation 
function
\begin{eqnarray}
g_{++}(r)=\frac{<\rho^+(0) \rho^+(r)>}{<\rho^+>^2}     & \hspace{2cm} &
g_{+-}(r)= \frac{<\rho^+(0) \rho^-(r)>}{<\rho^+> <\rho^->}
\end{eqnarray}
for the monopole-monopole (++) and monopole-antimonopole (+-) cases
($\rho^+$ and $\rho^-$ are the densities of timelike trajectories
wrapped in the positive and negative time direction at a fixed
time-slice). This quantity is the probability of finding a
(anti)monopole at distance $r$ from a given monopole normalized 
by the same probability in an uncorrelated situation. 
For the free case $g(r)=1$,
while $g(r)>1$ ($g(r)<1$) means that at a distance $r$ we have more
(less) particles then in the free case due to attraction
(repulsion). For a solid we expect a multiple peak structure
corresponding to several shells while in a liquid-like situation we
usually expect only one large bump. This is just the situation for
the monopole-antimonopole case as we see in fig.~\ref{grfig} (left),
from which a liquid-like behavior is qualitatively evident, with stronger interaction at high $T$ \cite{shuryakliquid},
where the bump gets higher. In fig.~\ref{grfig} (right) we see, from the depletion region in $g_{++}(r)$ at short
distances, that monopoles repel monopoles while at this distance there are more antimonopoles
than in the free case ($g(r)>1$), that means attraction between monopoles and antimonopoles.
Notice also the very good scaling of $g(r)$, which means that even the correlation function
is a well defined physical quantity.

Data have been fitted according to $g(r)=e^{-V(r)/T}$, 
where $V(r)=\alpha_M e^{-r/\lambda}/r$ is a screened Coulomb
potential, obtaining $\lambda \sim$ 0.1 fm. 
Further analysis~\cite{shuryakliquid} has led to an estimate of the 
plasma parameter
$\Gamma = \alpha_M (4\pi\rho/3 )^{1/3} / T$ ($\sim$
interaction/kinetic energy) between $2$ and $4.5$, which 
is a quantitative evidence for a liquid-like
(nontrivial) interaction.

\section{The gauge dependence problem}

The definition of thermal monopoles, being based on the Abelian
projection, is naturally subject to gauge dependence, or more precisely 
to dependence on the gauge where the Abelian projection is performed.
 That can be 
dramatic: for instance in Landau gauge wrapping
trajectories are practically absent~\cite{ourpaper} at any $T$.
More than that, even within the same Abelian projection 
Gribov copy ambiguities (due to the existence of
multiple local maxima for the gauge functional) 
may lead to systematic effects.
For example if we start
the maximization of the MAG functional from a Landau gauge fixed instead of
a random thermalized configuration, we find a different result
(fig.~\ref{densfig} (right)) and -above all- the scaling is lost.
Further investigations should be made by looking for the
global maximum of the MAG functional via better algorithms, like simulated annealing~\cite{bornya02}.

Regarding the dependence on the gauge chosen for Abelian projection,
one could try to look at different definitions of monopoles. One possibility
is the gauge independent definition given in Ref.~\cite{DTkondo}, where a partial gauge fixing 
$SU(2)_G\times SU(2)/U(1)_{G'}\rightarrow SU(2)_{G=G'}$ is done maximizing
\begin{equation}
F_{nMAG}=\sum_{x, \mu} \mbox{Tr}((\vec \sigma \cdot \vec n^{G'}(x))
U_\mu^G(x) (\vec \sigma \cdot \vec n^{G'}(x+\mu)) U_\mu^{G\dag}(x))
\end{equation}
The adjoint color field $\vec n$ (a unit vector) is linked to $U_\mu(x)$  via 
$V_\mu(x) \propto U_\mu(x) + (\vec \sigma \cdot \vec n(x)) 
U_\mu(x) (\vec \sigma \cdot \vec n(x+\mu))$,
where $V_\mu(x)$ is the unitarized $\vec n$-proportional part 
in the CFN \cite{DTkondo} Hodge decomposition of the gauge field $A_\mu$.
Monopoles are then defined by applying the usual De Grand - Toussaint
definition   on 
the phase of $SU(2)$ gauge invariant plaquettes
\begin{equation}
\theta_{\mu\nu}(x)=\mbox{arg}{\left[\mbox{Tr}\frac{1+\vec \sigma \cdot
      \vec n(x)}{\mbox{Tr}(1)} V_\mu(x) V_\nu(x+\mu) V_\mu^\dag(x+\nu) V_\nu^\dag(x)\right]}_{\mbox{mod}\ 2\pi}
\label{DTnMAG}
\end{equation}
As this definition does not depend on simultaneous $SU(2)$
transformations $G(x) (\vec \sigma \cdot \vec n(x)) G^\dag(x)$ and 
$G(x) V_\mu(x) G^\dag(x+\mu)$ we are free to choose $\vec \sigma \cdot \vec n(x)=\sigma_3$ everywhere: in this way 
$F_{nMAG}=F_{MAG}$ and $V_\mu(x)$ is the diagonal part of $U_\mu(x)$
selected in the same way as in the conventional Abelian projection $V_\mu(x)=U_\mu^{diag}=\mbox{diag}(U^1_1,U^2_2)$
(unitarized); finally the projector $\mbox{Tr}\left(\frac{1+\sigma_3}{\mbox{Tr}(1)} \ldots \right)$ in \ref{DTnMAG} takes 
out the diagonal phase of the plaquette $\exp(i \theta_{\mu\nu} \sigma_3)\rightarrow \exp(i \theta_{\mu\nu})$.
The prescription in Ref.~\cite{DTkondo} with nMAG applied to \ref{DTnMAG}
is then completely equivalent, by construction, to the usual Abelian monopole 
definition based on Abelian projection in the MAG gauge: therefore it
does not provide, at least for gauge group $SU(2)$, 
an independent definition useful for further studies. However
in this respect the nMAG formalism, based on theoretical evidences for 
Abelian dominace (see \cite{DTkondo} and references therein), can be regarded 
as a justification for adopting the MAG.

\section{Conclusions}
We presented a study of the density and correlation function
of thermal Abelian monopoles in the deconfined phase of $SU(2)$ pure
gauge theories: both of them show a good scaling to the continuum
and confirm the presence of non-trivial interactions
among monopoles which could be relevant to QGP properties.
Further analysis of the properties of wrapping trajectories is in
progress and will give access to other properties of 
thermal monopoles, such as their physical mass.

Dependence on the gauge chosen for Abelian projection as
well as Gribov copy effects
remain at present a problem
in the physical interpretation of thermal monopoles which needs 
clarification. Different gauge invariant definitions~\cite{DTkondo} 
lead to monopoles which are identical 
to MAG monopoles and are therefore of no benefit in this respect,
even if provide further inside in the MAG itself.

A different direction to be pursued in order
to prove the physical nature of thermal monopoles 
is to study the excess of non-Abelian action around
wrapped monopole trajectories~\cite{actionexcess}.

\end{document}